	\documentclass[11pt]{article}
	\usepackage{amsmath}
	\usepackage{graphicx,psfrag,epsf}
	\usepackage{enumerate}
	\usepackage{natbib}
	\usepackage{url} 
	
	\newcommand{\blind}{1}
	
\usepackage{color, mathtools,ulem, bbding}
\usepackage{adjustbox}
\usepackage{tikz}
\usetikzlibrary{shapes,arrows, fit}
\usetikzlibrary{positioning,calc}
\setcitestyle{citesep={;}, aysep={,}}

\newcommand{\bftau} {\mbox{\boldmath $\tau$}}

\begin{document}
	
\def\spacingset#1{\renewcommand{\baselinestretch}%
	{#1}\small\normalsize} \spacingset{1}

\ifpdf
	\DeclareGraphicsExtensions{.pdf, .png, .jpg, .tif}
	\else
	\DeclareGraphicsExtensions{.eps, .jpg}
	\fi

\if1\blind
{
	\title{\bf Bayesian Joint Modeling of Multiple Brain Functional Networks}
	\author{ Joshua Lukemire \\ Department of Biostatistics and Bioinformatics, Emory University, USA. \\
	Suprateek Kundu \\ Department of Biostatistics and Bioinformatics, Emory University, USA. \\
		Giuseppe Pagnoni\thanks{This work was supported by the European Community Marie Curie Action IRG (Call FP7-PEOPLE-2009-RG, project 249329 ``NBC-EFFORT").}\\  Department of Biomedical, Metabolic and Neural Sciences,\\
		University of Modena and Reggio Emilia, Italy.\\
		and\\
		Ying Guo \thanks{Research reported in this publication was supported by the National Institute Of Mental Health of the National Institutes of Health under Award Number RO1 MH105561 and R01MH079448. The content is solely the responsibility of the authors and does not necessarily represent the official views of the National Institutes of Health.} \\
		Department of Biostatistics and Bioinformatics, Emory University, USA.
		}
	\date{}
	\maketitle
} \fi

\if0\blind
{
	\bigskip
	\bigskip
	\bigskip
	\begin{center}
		{\LARGE\bf Bayesian Joint Modeling of Multiple Brain Functional Networks}
	\end{center}
	\medskip
} \fi


\begin{abstract}
Brain function is organized in coordinated modes of spatio-temporal activity (functional networks) exhibiting an intrinsic baseline structure with variations under different experimental conditions. Existing approaches for uncovering such network structures typically do not explicitly model shared and differential patterns across networks, thus potentially reducing the detection power. We develop an integrative modeling approach for jointly modeling multiple brain networks across experimental conditions. The proposed Bayesian Joint Network Learning approach develops flexible priors on the edge probabilities involving a common intrinsic baseline structure and differential effects specific to individual networks. Conditional on these edge probabilities, connection strengths are modeled under a Bayesian spike and slab prior on the off-diagonal elements of the inverse covariance matrix. The model is fit under a posterior computation scheme based on Markov chain Monte Carlo. Numerical simulations illustrate that the proposed joint modeling approach has increased power to detect true differential edges while providing adequate control on false positives and achieving greater accuracy in the estimation of edge strengths compared to existing methods. An application of the method to fMRI Stroop task data provides unique insights into brain network alterations between cognitive conditions which existing graphical modeling techniques failed to reveal.
\end{abstract}

\noindent%
\vfill

\spacingset{1} 
\section{Introduction}

The study of the neural bases of human cognition has made rapid progress with the advent of modern brain imaging techniques. Among these, functional Magnetic Resonance Imaging (fMRI) has allowed researchers to associate specific sets of brain regions with the performance of a variety of cognitive, sensory, and motor tasks by detecting activation-related local changes in the blood-oxygen-level-dependent (BOLD) signal. In recent years, the interest of the neuroimaging community has somewhat shifted from localized brain activation to functional connectivity, that is, how different regions of the brain change their activity together as a functionally coherent circuit (see Smith et al., 2011, for a review). 


Research on functional connectivity involves investigations of resting state fMRI and task-based fMRI seeking to identify coherent patterns of spatio-temporal activity during rest and task, respectively \citep{Biswal1995}. Functional connectivity is most often assessed in terms of brain networks, i.e. sets of connections between different brain regions under a graph-theoretic approach, where each connection represents a path of information transmission. Several approaches have been proposed for modeling the brain network in terms of a graph; these include pairwise and partial correlation analysis  \citep{Salvador2005, wang2016} and sparse inverse covariance or precision matrix estimation \citep{wang2016}. \cite{Smith2011} showed the precision matrix approach and the related partial correlations method to be very successful in distinguishing a true, direct functional connection between two nodes from an apparent one mediated by a third common cause.

Many researchers are interested in comparing brain networks across cognitive states induced by experimental conditions with the aim of identifying functional connections whose strengths reflect differences or commonalities between conditions \citep{ Fox2007}. Under a graph-theoretic approach, edges featuring differential strengths correspond to brain connections that are more activated or suppressed during one experimental condition as compared to others. Such edges are potentially of great clinical and translational significance. On the other hand, connections shared across experimental conditions may represent an {\it intrinsic} functional network architecture which is common across varying cognitive states \citep{Fox2007}. The comparison of brain networks across multiple conditions may be performed on a single subject or, as in our case, at a group level \citep{Smith2011}, with the latter being able to average out subject-specific idiosyncrasies and potentially providing greater power to detect underlying biological differences and similarities \citep{kim2015}.

Existing approaches for comparing multiple brain networks typically estimate each brain network separately and then use mass-univariate hypothesis testing to infer significant differences between these networks while controlling the family-wise error rate \citep{genovese2002}. These approaches, although valuable, may have reduced power to detect true differences \citep{fornito2013} due to the fact that they have to adjust for a massive number of multiple comparisons, and because they do not borrow information across networks, which results in less accurate estimates. Alternatively, network metrics and statistics can be used to avoid testing every possible connection \citep{fornito2013}. These techniques improve statistical power but come at the cost of reduced explanatory value since they typically do not provide inferences at the level of individual connections.

Recently, there has been a limited growth in the development of penalized approaches for the joint estimation of multiple graphical models. These approaches \citep{Guo2011,Danaher2014,Zhu2014} rely on penalization to enforce sparsity and typically smooth over the strength of connections across networks to enforce shared edges, which is a useful modeling assumption but may not be supported in practical brain network applications. With the exception of a recent work by \cite{Belilovsky2016}, who developed a penalized neighborhood selection approach to obtain point estimates for brain networks, very few existing penalized methods have been vetted for estimating multiple brain networks, to our knowledge. Unfortunately, the approach by \cite{Belilovsky2016}  cannot be used to obtain positive definite precision matrices, precluding accurate quantification of edge strengths in terms of partial correlations. Moreover, the above penalized approaches only report point estimates; they do not provide measures of uncertainty for the brain networks, which are often desirable in accounting for the underlying heterogeneity in a group level analyses, as well as quantifying estimation errors in brain imaging studies.  \cite{kim2015} suggested that comparing brain networks based on penalized approaches may result in misleading inferences since the estimated network differences may be artifacts resulting from estimation errors under point estimates. 

On the other hand, a number of Bayesian spike and slab approaches \citep{yu2016, peterson2015}, and continuous shrinkage methods \citep{carvalho2010, polson2010, piironen2017, li2017} have been proposed for precision matrix estimation in recent years, with some applications to brain network analysis \citep{Mumford2014}. Though Bayesian approaches have proven extremely useful in estimating individual brain networks, few attempts have been made to develop Bayesian methods for the joint estimation of multiple graphical networks. Some existing Bayesian methods for jointly estimating multiple graphs include the approach by \cite{yajima2012}, who focused on multiple directed acyclic graphs, and the Bayesian Markov random field approach by \cite{peterson2015} for estimating multiple protein-protein interaction networks. The former cannot be used to obtain undirected brain networks which is the focus of this article, and the latter is only applicable to examples involving a small number of nodes. We note that there is some recent work on jointly estimating multiple temporally dependent brain networks \citep{Qiu2016, lin2017}, but these approaches cannot be directly generalized for the integrative analysis of multiple brain networks across different experimental conditions or cohorts. The above discussion suggests a clear need for developing flexible Bayesian approaches for joint estimation of multiple brain networks which pool information across graphs to provide more accurate inferences. 

In this article, we develop a Bayesian Gaussian graphical modeling approach for estimating multiple networks. This approach models the probability of a connection as a parametric function of a baseline component shared across networks and differential components unique to each network. The shared and differential effects are modeled under a Dirichlet process (DP) mixture of Gaussians prior \citep{muller1996}, and the edge probabilities are estimated by pooling information across experimental conditions, thereby resulting in the joint estimation of multiple brain networks. The role of the edge probabilities is twofold - they characterize uncertainty in network estimation and enable direct testing of shared and differential patterns across networks after multiplicity corrections. The connection strengths are encapsulated via network specific precision matrices, which are modeled separately for each network under a spike and slab Bayesian graphical lasso prior informed by the above edge probabilities. Adopting a joint modeling approach involving a combination of a parametric link function with flexible DP priors on the components results in an interpretable and flexible method that enables more accurate estimation of edge strengths and provides improved power to detect true differential connections, while ensuring adequate control for false positives, as demonstrated via extensive numerical experiments. Another important advantage in using the DP prior on the components is the robustness to the specification of the parametric link function, as evident from the results in Table \ref{probitTable}. The approach, denoted as Bayesian Joint Network Learning (BJNL), is implemented via a fully Gibbs posterior computation scheme which proceeds via Markov chain Monte Carlo (MCMC). 

Our method was applied to a fMRI Stroop task experiment  \citep{stroop1935, khachouf2017} in which data were collected under blocks of passive fixation and blocks of task performance, requiring the participants to alternately execute the same task with a maximum or minimum degree of voluntary effort investment. An exploratory analysis of the data, which involved deriving the subject-specific network for each of the 45 subjects under the task and rest conditions using the graphical lasso \citep{Friedman2008}, and then estimating the group level probability for each edge by combining the edge sets across all subjects, followed by a $K$-means algorithm on the edge probabilities, revealed clearly defined and well separated clusters for these probabilities. This provides a strong motivation for a DP mixture approach to cluster the edge probabilities. We sought to test the ability of the method to assess varying degrees of network differences between passive fixation and task performance, as well as between effortful and relaxed task performance, while also estimating the commonalities across the conditions. Results from BJNL provide insights into how the brain network reorganizes under different cognitive conditions. Specifically, we found that brain connections are dramatically different when subjects are actively engaged in the task as compared with the rest condition. In addition to the identification of a considerable number of connections that were altered under the task vs. passive fixation, we found the  topological characteristics of the brain networks to be different. For example, the rest condition was found to be associated with more efficient information transmission. On the other hand, a comparison of brain networks under the two task conditions corresponding to a varying degree of cognitive exertion revealed similar topological features but also demonstrated alterations in brain networks that are associated with high level cognitive processing. In addition, our approach characterized the uncertainty inherent in the group level analysis in terms of edge probabilities and posterior distributions for different graph metrics. In contrast, alternate analyses involving separate estimation of multiple brain networks using the graphical lasso  and joint graphical lasso \citep{Danaher2014} approaches, followed by permutation testing, revealed feeble or no connectivity differences, as elaborated in Section 5. 

This paper is organized as follows. In Section 2 we describe the fMRI data set and illustrate the proposed approach. Section 3 provides details about the posterior computation strategy for implementing the BJNL. Section 4 reports extensive numerical studies comparing the proposed BJNL to competing approaches. Section 5 discusses the results of the application of the method to the fMRI data set, and Section 6 summarizes our findings. The Supplementary Materials contain explicit posterior computation steps, details of the fMRI preprocessing, and additional numerical and Stroop task results. For ease of use and as a starting point for further extension of the method, we provide a Matlab GUI implementation of our method in the Supplementary Materials as well.

\section{Methodology}

\subsection{Description of the fMRI data set}

\begin{figure}
	\begin{center}
		\includegraphics[width=1\linewidth]{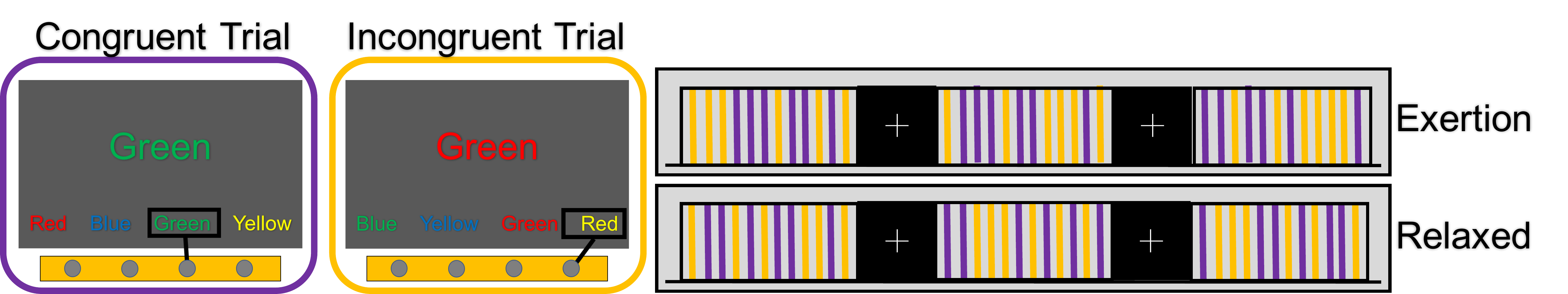} 
		\caption{\footnotesize An illustration of the Stroop task involving task blocks of congruent and incongruent trials, indicated by purple bars and yellow bars respectively, and fixation blocks denoted by a centrally fixated cross. The purple and yellow bars are expanded into two boxes, and the correct button presses are indicated with a rectangle within each box. Subjects were instructed to perform odd-numbered runs ``with maximum exertion'' (EXR condition) and even-numbered runs ``as relaxed as possible'' (RLX condition).}
		\label{Stroop}
	\end{center}
\end{figure}

Forty-five volunteers participated in the study. All subjects were right handed with an average age of 21.9 (SD = 2.2) years. MRI scanning was performed at the N.O.C.S.A.E Hospital in Baggiovara (MO), Italy, using a 3T Philips Achieva scanner. For each subject, the imaging session consisted of the collection of 6 echo-planar imaging (EPI) runs (112 volumes each, TR=2.5s, 25 axial slice, $3 \times 3 \times 3$ mm voxels) and a T1-weighted high-resolution volume (180 sagittal slices, 1mm isotropic voxels) for anatomical reference. While in the scanner, subjects performed a 4-color version of the Stroop task with a button-press response modality \citep{gianaros2005}. In this task, subjects are presented with a color word displayed in colored fonts in the center of a computer screen and are asked to press a button on a response device corresponding to the font color of the stimulus. There are two types of trials: \emph{congruent} trials, where the font color matches the text (e.g., the word `RED' in red fonts), and \emph{incongruent} trials, where the font color does not match the text (e.g., the word `RED' in green fonts). The `Stroop effect' refers to a significant slowing of response times to the incongruent trials compared to the congruent ones \citep{stroop1935}. Figure \ref{Stroop} illustrates the Stroop task experiment.  

Stimuli were presented in (task) blocks of 30s containing 6 congruent and 6 incongruent trials appearing in a pseudo-random order with a 2.5s inter-trial interval. Each task block was alternated with 25s-blocks of passive fixation on a centrally presented cross. Six fMRI runs were collected for each subject, with each run consisting of 4 blocks of task and 5 blocks of passive fixation appearing in ABABABABA order (A=passive fixation, B=task). Crucially, subjects were instructed to perform odd-numbered runs ``with maximum exertion'' (EXR condition) and even-numbered runs ``as relaxed as possible'' (RLX condition). This scheme was reversed for a subset of volunteers to check for potential order effects.  A major aim of the study was to compare the brain connectivity under REST (passive fixation) and the two TASK conditions (RLX and EXR) \citep{khachouf2017}.

\subsection{Bayesian modeling of multiple networks}

We develop a novel Bayesian approach for jointly estimating multiple group-level brain functional networks from multi-subject fMRI data. For each subject, the data are demeaned and pre-whitened across time points, where the pre-whitened fMRI observations are considered statistically independent. The pre-whitened fMRI data over $p$ nodes or regions of interest (ROI) for the $i$-th subject and $g$th experimental condition at time point $t$ is denoted by ${\bf y}_{it}(g)=(y_{it1}(g),\ldots,y_{itp}(g)), i=1,\ldots, n, t=1,\ldots,T_{ig}, g=1,\ldots,G$. Our goal is to jointly estimate multiple networks denoted by $\mathcal{G}_1,\ldots,\mathcal{G}_G$ using Gaussian graphical models characterized by sparse inverse covariance matrices. The graph $\mathcal{G}_g$ is defined by the vertex set $\mathcal{V}=\{1,\ldots,p \}$ containing $p$ nodes and the edge set $\mathcal{E}_g$ containing all edges/connections in the graph $\mathcal{G}_g,g=1,\ldots,G$.

The pre-whitened fMRI measurements for $g$-th experimental condition are modeled as $ {\bf y}_{it}(g) \sim N_p({\bf 0},\boldsymbol{\Omega}^{-1}_g)$, $ i=1,\ldots,n, t=1,\ldots,T_{ig},g=1,\ldots,G,$ where 
\begin{gather}
\scalebox{0.8}{$
   \begin{aligned}
	\pi(\boldsymbol{\Omega}_g) = C^{-1}_g\prod_{k=1}^p  E(\omega_{g,kk}; \frac{\alpha}{2}) \bigg\{\prod_{k<l} w_{g,kl} N(\omega_{g,kl};0,\tau^{-1}_{g,kl}) + (1-w_{g,kl}) DE( \omega_{g,kl}; \lambda_0) \bigg\} I(\boldsymbol{\Omega}_{g}\in M^{+}),  \label{eq:base}
   \end{aligned}  $}
\end{gather}

where $\pi(\cdot)$ denotes the prior distribution, $\omega_{g,kl}$ and $w_{g,kl}$ denote the strength and probability of the functional connection between nodes $k$ and $l$ for network $\mathcal{G}_g$ respectively, $M^{+}$ denotes the space of all positive definite matrices, $I(\cdot)$ denotes the indicator function, $C_g$ is the intractable normalizing constant for the prior on the precision matrix, $N_p(\cdot;{\bf 0},\boldsymbol{\Sigma})$ denotes a $p$-variate Gaussian distribution with mean ${\bf 0}$ and covariance $\boldsymbol{\Sigma}$, and $E(\alpha)$ and $DE(\lambda)$  denote the exponential and double exponential distributions with scale parameters $\alpha^{-1}$ and $\lambda^{-1}$ respectively. Small values of the scale parameters $\tau_{g,kl} \sim \pi(\tau_{g,kl})$ and $\lambda_0^{-1}$ in equation (3)  result in a spike and slab prior \citep{george1993} on the precision off-diagonals, so that $\boldsymbol{\Omega}_g \sim \pi(\boldsymbol{\Omega}_g)$ is denoted as the {\it spike and slab Bayesian graphical lasso}. The spike and slab prior shrinks the values corresponding to absent edges toward zero and encourages values away from zero for important connections. 
The slab component is modeled under a Gaussian distribution having thick tails under small values of the precision parameter, while the spike component is modeled under a double exponential distribution having a sharp spike at zero under a large value of $\lambda_0$. It is straightforward to show that $C_g<\infty$ so that the prior in model (\ref{eq:base}) is proper using the results in \cite{wang2012}.

\uline{\noindent \it {Pooling Information Across Experimental Conditions:}} Information is pooled across experimental conditions to estimate the edge weights $w_{g,kl}$, $k\ne l,k,l=1,\ldots,p,$ leading to joint estimation of multiple networks. Note that by pooling information to model the edge probabilities instead of the edge strengths, we are able to jointly model multiple brain networks without constraining the edge strengths in separate networks to be similar. The prior weights represent the unknown probabilities of having functional connections, and are modeled via a parametric link function comprising unknown shared and differential effects as described below
\begin{gather}
\scalebox{1}{$
   \begin{aligned}
	w_{g,kl} = h(\eta_{0,kl},\eta_{g,kl}), \mbox{ } \eta_{0,kl}\sim f_0, \mbox{ }\eta_{g,kl} \sim f_g, \mbox{ } f_0\sim DP(MP_0), \mbox{ } f_g\sim DP(MP_0),
	\label{eq:weights}
   \end{aligned}  $}
\end{gather}

for $k\neq l, k,l=1,\ldots,p, g=1,\ldots,G,$ where $h(\cdot)$ is the parametric link function relating the probability for edge $(k,l)$ in network $\mathcal{G}_g$ to the network specific differential effect ($ \eta_{g,kl} $) and common effect ($ \eta_{0,kl} $) across all networks, and $DP(MP_0)$ denotes a Dirichlet process mixture prior defined by the precision parameter $M$ and base measure $P_0 \equiv N(0,\sigma^2_{\eta})$. The Dirichlet process mixture prior induces a flexible class of distributions on the edge probabilities and also results in clusters of edges having the same prior inclusion probabilities, enforcing parsimony in the number of model parameters. The number of clusters and the cluster sizes are unknown and controlled via the precision parameter $M$  \citep{antoniak1974}. 

Under specification (\ref{eq:weights}), the baseline effect $\eta_{0,kl}$  represent the shared feature for edge $(k,l)$ which is estimated by pooling information across experimental conditions, resulting in the joint estimation of multiple networks. The baseline effect controls the overall probability of having an edge across all networks, while the differential effects contribute to the network specific variations which are estimated using the information from individual experimental conditions. For example, large differences between $\eta_{g,kl}$ and $\eta_{g',kl},g\ne g'$ potentially imply a differential status for edge $(k,l)$ between $\mathcal{G}_g$ and $\mathcal{G}_{g'}$. On the other hand when $\eta_{g,kl}=\eta_{g',kl},g\ne g'$, the model specifies equal probability for edge $(k,l)$ in networks $\mathcal{G}_g$ and $\mathcal{G}_{g'}$. For ease in interpretability we choose a logistic form link in (\ref{eq:weights}) as  $	h(\eta_{0,kl}, \eta_{g,kl}) =\exp\{\eta_{0,kl} + \eta_{g,kl} \}/ [ 1 + \exp\{\eta_{0,kl} + \eta_{g,kl} \} ], g=1,\ldots,G,$ so that $\eta_{0,kl}+\eta_{g,kl}$ can be interpreted as the log odds of having the edge $(k,l)$ in the network $\mathcal{G}_g$, and the log odds ratio of having edge $(k,l)$ in the brain network $\mathcal{G}_g$ versus $\mathcal{G}_{g'}$ can be expressed as $ \eta_{g,kl} - \eta_{g',kl}$ ($g \ne g'$). A schematic representation of the proposed model is illustrated in Figure \ref{parameterChart}.

Note that the parameters $ \eta_{0,kl}, \eta_{g,kl},$ in  (\ref{eq:weights})  are not identifiable since $	h(\eta_{0,kl}, \eta_{g,kl}) = h(\eta_{0,kl}+ c, \eta_{g,kl}-c) $ for any real constant $c$. However, the functionals of interest such as the log odds ($ \eta_{0,kl}+\eta_{g,kl}$), the log-odds ratio ($ \eta_{g,kl}-\eta_{g',kl}$), and the edge probabilities themselves are clearly identifiable, which is adequate for our purposes. The proposed specification (\ref{eq:weights}) is purposely overcomplete, which is an issue routinely arising in Bayesian models. By ``overcomplete," we mean that we include $G+1$ parameters in the weights model when $G$ parameters would suffice. Such overcompleteness allows us to pool information in a systematic manner, and ensures computational efficiency and interpretability in terms of shared and differential group effects and is designed to avoid any problems in identifiability of functionals of interest  - refer, for example, to  \cite{ghosh2009}.

Our treatment of the edge weights is motivated by existing literature on modeling binary or ordered categorical responses using mixture distributions \citep{kottas2005, jara2007, gill2009, canale2011}. Specifically we are able to achieve both the interpretability discussed above and a high degree of flexibility while also reducing the sensitivity to the link function and enabling straightforward posterior computation. A similar approach was taken by \cite{durante2017} who modeled structural connections in a population of networks via a mixture of Bernoulli distributions, although they did not focus on joint estimation of multiple networks.

\tikzstyle{recBlock} = [rectangle, draw, fill=white!10, text centered, rounded corners, minimum size=1cm]
\tikzstyle{edgeWeightsBackground} = [rectangle, draw, fill=yellow!5, text centered, rounded corners, minimum size=5cm, minimum height=6cm]
\tikzstyle{dirPBackground} = [rectangle, draw, fill=yellow!5, text centered, rounded corners, minimum size=2.1cm, minimum height=6cm]
\tikzstyle{likelihoodBackground} = [rectangle, draw, fill=green!5, text centered, rounded corners, minimum size=1.5cm, minimum height=6cm]
\tikzstyle{dirpblock} = [rectangle, draw, fill=blue!10, text centered, rounded corners, minimum size=2cm]
\tikzstyle{ellBlock} = [draw, ellipse, minimum size=1cm, fill=white!10]
\tikzstyle{circleBlock} = [draw, circle, minimum size=1cm, fill=white!10]
\tikzstyle{circleBlockClear} = [draw, circle, minimum size=1cm, fill=white!10, opacity=0, line width=0]
\tikzstyle{diamondBlock} = [draw, diamond, minimum size=.5cm, fill=white!10]
\tikzstyle{line} = [draw, -latex']
\tikzstyle{edgeEstBackground} = [rectangle, draw, fill=yellow!2, text centered, rounded corners, minimum size=11.0cm, minimum height=8cm]
\tikzstyle{edgeStrengthBackground} = [rectangle, draw, fill=green!2, text centered, rounded corners, minimum size=2cm, minimum height=8cm]
\tikzstyle{hyperBackground} = [rectangle, draw, fill=yellow!5, text centered, rounded corners, minimum size=1.7cm, minimum height=6cm]

\begin{figure}[ht!]
	\begin{center}
		\scalebox{.57}{ 	\begin{tikzpicture}[node distance = 1cm, auto]
	\node [circleBlockClear] (anchorBlock0) {tt};
	\node [circleBlockClear, right=3cm of anchorBlock0] (anchorBlock) {t};
	
	\node [edgeEstBackground, left=-8.1cm of anchorBlock](edgeBackground){};
	\node [edgeStrengthBackground, right = 7.5cm of anchorBlock](edgeStrBackground){};
	
	\node [edgeWeightsBackground, right = -1.3cm of anchorBlock] (eWBB) {};
	\node [likelihoodBackground, right = 4.00cm of eWBB] (lBB) {};
	\node [dirPBackground, left = 0.30cm of eWBB] (DIRPB) {};
	\node [hyperBackground, right = 0.5cm of eWBB] (hyB) {};
	
	\node [recBlock, right=3cm of anchorBlock0] (etaTerm0) {$\eta_0$};
	\node [recBlock, above=1cm of etaTerm0] (etaTerm1) {$\eta_1$};
	\node [recBlock, below=1cm of etaTerm0] (etaTerm2) {$\eta_2$};
	
	\node [recBlock, above right=0.00cm and 0.7cm of etaTerm0] (wTerm1) {$w_1 = h(\eta_0, \eta_1)$};
	\node [recBlock, below right=0.00cm and 0.7cm of etaTerm0] (wTerm2) {$w_2= h(\eta_0, \eta_2)$};
	
	\node [diamondBlock, above right=0.35 cm and 4.62cm of wTerm1] (omegaTerm1) {$\boldsymbol{\Omega}_1$};
	\node [diamondBlock, below right=0.35cm and 4.62cm of wTerm2] (omegaTerm2) {$\boldsymbol{\Omega}_2$};
	
	\node [recBlock, left=0.7cm of etaTerm0] (pnot) {$P_0, M, \sigma_{\eta}^2$};
	
	\node [circleBlock, below=0.4cm of omegaTerm1] (Y1) {$\boldsymbol{Y}_1$};
	\node [circleBlock, above=0.4cm of omegaTerm2] (Y2) {$\boldsymbol{Y}_2$};
	
	\node [recBlock, below left=1.5cm and 2.45cm of omegaTerm1] (hyper1) {$\alpha, \lambda_0$};
	\node [recBlock,  above left = -0.75 and 2.5cm of omegaTerm1] (tau1) {$\tau_1^{-1}$};
	\node [recBlock,  below left = -0.75 and 2.5cm of omegaTerm2] (tau2) {$\tau_2^{-1}$};
	
	\path [line] (etaTerm0) -- (wTerm1);
	\path [line] (etaTerm1) -- (wTerm1);
	\path [line] (etaTerm0) --(wTerm2);
	\path [line] (etaTerm2) --  (wTerm2);
	
	\path [line] (wTerm1) -- (omegaTerm1);
	\path [line] (wTerm2) -- (omegaTerm2);
	\path [line] (pnot) -- (etaTerm0);
	\path [line] (pnot) -- (etaTerm1);
	\path [line] (pnot) -- (etaTerm2);
	\path [line] (omegaTerm1) -- (Y1);
	\path [line] (omegaTerm2) -- (Y2);
	\path [line] (hyper1) -- (omegaTerm1);
	\path [line] (hyper1) -- (omegaTerm2);
	\path [line] (tau1) -- (omegaTerm1);
	\path [line] (tau2) -- (omegaTerm2);
	
	\node[text width=2cm,  below right = 0.01cm and -1.5cm of lBB]  {\scriptsize{Likelihood}};
	\node[text width=3cm,  below right = 0.01cm and -3.7cm of eWBB]  {\scriptsize{Edge Weights}};
	\node[text width=2cm,  below = 0.01cm of DIRPB]  {\scriptsize{Dir. Process} };
	\node[text width=2cm,  below = 0.01cm of hyB]  {\scriptsize{Hyperparameters} };
	\node[text width=5cm,  above right = 0cm and -7.2cm of edgeBackground]  {Spike and Slab Prior};	
	\node[text width=5cm,  above right = 0cm and -2.5cm of edgeStrBackground]  {Observed Data};	
	
	\end{tikzpicture} }
	\end{center}
	\caption{\footnotesize Directed graph illustrating the relationships between the model parameters for the  case of two experimental conditions represented by fMRI data matrices $\boldsymbol{Y}_1$ and $\boldsymbol{Y}_2$. Rectangular nodes correspond to parameters which are updated or tuned, diamond-shaped nodes correspond to parameters involved in the likelihood, and the circular nodes correspond to the observed data.}
	\label{parameterChart}
\end{figure}
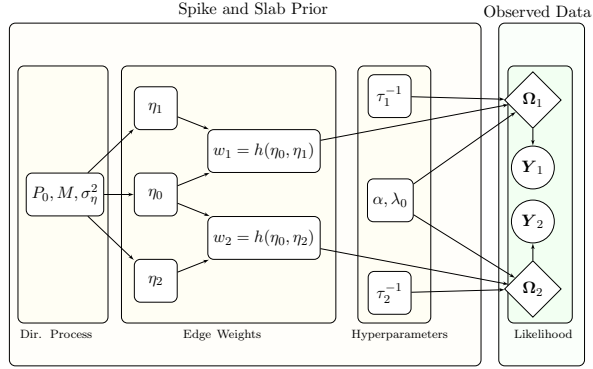

\section{Posterior Computation}
We design a block Gibbs sampler in order to fit the proposed model (\ref{eq:base}). The sampler enables data adaptive shrinkage by introducing latent scale parameters to sample the precision matrix off-diagonals corresponding to the spike component under a scale mixture representation of Gaussians while defining conjugate priors on the precision parameters in the slab component. Define edge inclusion indicators as $\delta_{g,kl}=1$ if edge $(k,l)$ is included in $\mathcal{G}_g$, and $\delta_{g,kl}=0$ otherwise, where $ P(\delta_{g,kl}=1) = w_{g,kl}$. The augmented likelihood for equation (\ref{eq:base}) can be written as 
 

{\scriptsize
\begin{flalign}
   \label{eq:px}
\pi(\boldsymbol{\Omega}_g \mid \lambda_0,\bftau_g, \bftau^*_g) &= C^{-1}_{\tau,g} 1(\boldsymbol{\Omega}_g \in M^{+}) \prod_{l=1}^p Exp(\omega_{g,ll}; \alpha/2) \times \prod_{l=1}^p\prod_{k<l}  w_{g,kl}^{\delta_{g,kl}}(1-w_{g,kl})^{1-\delta_{g,kl}} \nonumber  \\
&\times \prod_{l=1}^p\prod_{k<l}  [N(\omega_{g,kl}; 0, \tau^{-1}_{g,kl})]^{\delta_{g,kl}} [\int N(\omega_{g,kl}; 0, \tau^{-1}_{g,kl})Exp(\tau_{g,kl};\frac{\lambda_0^2}{2}) d\tau_{g,kl}]^{1-\delta_{g,kl}}, \mbox{ with }  \nonumber \\
 \pi(\bftau_{g}, \bftau^*_{g}) &\propto C_{\tau,g} \bigg(\prod_{l=1}^p \prod_{k<l} Ga(\tau_{g,kl}; a_\tau, b_\tau) \times Exp(\tau^*_{g,kl};\lambda^2_0/2) \bigg),
\end{flalign}
}

where $ \bftau_{g}=\{ \tau_{g,kl}, k\ne l, k,l=1,\ldots,p\}$, $ \bftau^*_{g}=\{\tau^*_{g,kl}, k\ne l, k,l =1,\ldots,p \}$, $Ga(\cdot;a_\tau,b_\tau)$ corresponds to a Gamma distribution with mean $a_\tau/b_\tau$, and $ C_{\tau,g}$ is the intractable normalizing constant which cancels out in the expression for $ \pi(\boldsymbol{\Omega}_g, \lambda_0,\bftau_g, \bftau^*_g)$ to yield a marginal prior $ \pi(\boldsymbol{\Omega}_g, \lambda_0,\bftau_g)$ as in  (\ref{eq:base}) after integrating out $\bftau^*_{g}$. In our implementation we pre-specify $\lambda_0=100$ to ensure a sharp spike at zero leading to strong shrinkage for precision off-diagonals corresponding to absent edges. On the other hand, we choose  $a_\tau$ and $b_\tau$ such that $a_\tau/b_{\tau}$ is small, enabling adaptive thick tails for the Gamma prior on the latent scale parameters corresponding to the slab component.

We choose a logistic link function in (\ref{eq:weights}) for our purposes, although more general link functions can also be used. For implementing a fully Gibbs sampler, we rely on an approximation to the logistic function using a probit link, which employs a data augmentation scheme as in \cite{OBrien2004}. In particular, 
\begin{gather}
\scalebox{1}{$
   \begin{aligned}
    \frac{e^{\mu*}}{(1 + e^{\mu*})}  \approx \int_0^\infty {\bf t} \big( u; \mu*, \frac{\pi^2 (\phi-2)}{3 \phi} \big) du = \int_0^\infty N(u; \mu*, \frac{\pi^2(\phi - 2)}{3 \phi} \sigma_{\phi}^2) \pi(\sigma_{\phi}^2; \frac{\phi}{2}, \frac{\phi}{2}) du, \nonumber
   \end{aligned}  $}
\end{gather}
 where ${\bf t}(\cdot)$ denotes a t-distribution, $\pi(\sigma^2_\phi)$ corresponds to a inverse Gamma distribution, $\phi = 7.3$, and $u$ is the Gaussian latent variable used for data augmentation. This approximation results in sampling from a posterior that is approximately equal to the posterior under specification (\ref{eq:base})-(\ref{eq:weights}) using a logistic link function. Although such an approximation is used, we note that the resulting posterior computation is fully Gibbs since all MCMC samples are drawn from exact posterior distributions. Alternatively, one could adapt the Polya-gamma data augmentation in \cite{polson2013} for Bayesian logistic regression. However, the approximation in \cite{OBrien2004} works reasonably well in a wide variety of numerical studies in our experience.

Moreover, the stick-breaking representation \citep{sethuraman1994} is used for the Dirichlet process mixture prior in (\ref{eq:weights}), which facilitates posterior computation and can be written as 

\begin{gather}
\scalebox{0.8}{$
   \begin{aligned}
	\eta_{g, kl} \sim f_{g}, f_{g} = \sum_{h=1}^{\infty} \nu_{g, h} \delta_{ \eta_{g,h}^* }, \eta_{g,h}^* \sim N(0, \sigma_{\eta}^2), \nu_{g, h} = (v_{g,h} \prod_{l < h} [1 - v_{g, l}]), v_{g, h} \sim Beta(1, M), g = 0, \ldots, G, \label{eq:stickbreak}
   \end{aligned}  $}
\end{gather}

where $Be(\cdot)$ represents a Beta distribution. The slice sampling technique \citep{Walker2007} is used to sample the atoms from the infinite mixture in (\ref{eq:weights}), which significantly expedites computation. A step-by-step description of the posterior computation is provided in the Supplementary Materials.

\uline{\noindent \it{Edge Detection:}} The important edges in the graph under the proposed approach can be estimated by either including edges with high marginal inclusion probabilities or those with non-negligible absolute values for the precision off-diagonal elements, lying above a chosen threshold. We propose a strategy to choose such thresholds in a manner which controls the false discovery rate. Denoting $\zeta_{g,kl}$ as the marginal posterior exclusion probability for edge $(k,l)$ in network $\mathcal{G}_g$, one can compute the false discovery rate (FDR) as in \cite{newton2004} or \cite{peterson2015} as

\begin{gather}
\scalebox{1}{$
   \begin{aligned}
 FDR = \frac{\sum_{g=1}^G \sum_{k<l} \zeta_{g,kl}1(\zeta_{g,kl} < \kappa)}{\sum_{g=1}^G \sum_{k<l}1(\zeta_{g,kl} < \kappa)}, \mbox{ or } 
FDR = \frac{\sum_{g=1}^G \sum_{k<l} \zeta_{g,kl}1(|\hat{\omega}_{g,kl}| > \kappa^*)}{\sum_{g=1}^G \sum_{k<l}1(|\hat{\omega}_{g,kl}| > \kappa^*)}, \label{eq:fdr}
   \end{aligned}  $}
\end{gather}

depending on whether the edges are included based on posterior inclusion probabilities or edge strengths. Clearly the FDR increases with $\kappa/\kappa^*$, and one can choose a suitable threshold to control the FDR. In our numerical experiments we   found   that  choosing   the   edges   based   on whether   the   absolute  precision off-diagonals were greater than 0.1 results in overall   good   numerical performance and FDR values which are less than $0.03$ across a wide spectrum of scenarios. Hence we recommend this as a default threshold for estimating the network under our approach, and we note that the corresponding threshold for posterior probability for edge selection can be obtained as one which yields similar FDR as computed using expression (\ref{eq:fdr}). Once all the networks have been estimated using this strategy, the differential edges are identified as those which show up in one network but not the others. 

The proposed BJNL also provides a natural framework for testing differences in edge strengths between experimental conditions. As opposed to penalized likelihood approaches, which can only provide point estimates for the partial correlations, the MCMC samples from BJNL can be used to obtain the posterior distribution for differences in partial correlations. These differences can be Z-transformed using Fisher's method to obtain normally distributed values, which can then be tested for significance using T-tests after controlling for false discoveries \citep{benjamini1995}. In our experience based on extensive numerical studies we find that the differential edges based on the FDR-corrected criteria and the partial correlation based approach are overwhelmingly common; however, there could be some additional differences detected under the T-test due to significant variations in edge strengths for important edges.

\section{Numerical Studies}

\subsection{Simulation Setup}
We conducted a series of simulations to compare group level network estimation between BJNL and competing methods. These approaches include the graphical horseshoe estimator (HS) \citep{carvalho2010, li2017} which extends the horseshoe prior in regression settings to graphical model estimation, and the graphical lasso approach (GL) \citep{Friedman2008} which imposes $L_1$ penalty on the off-diagonals to impose sparsity, as well as the Joint Graphical Lasso (JGL) \citep{Danaher2014} which uses a fused lasso penalty to pool information across graphs while encouraging sparsity via a $L_1$ penalty. While both the HS and GL approaches estimate individual networks separately, the JGL approach is designed to jointly estimate multiple networks. The HS was implemented using Matlab codes provided on the author's website. The JGL and the graphical lasso were implemented using the {\it JGL} and {\it glasso} packages in R, respectively. Our method was implemented in Matlab, version 8.3.0.532 (R2014a), and a GUI implementing the method has been submitted as a Supplemental Material. 

The data for the simulation study was generated under a Gaussian graphical model for $n$=60 subjects with $T$=300 time points each and for dimensions $p=40,100$. Each subject had data corresponding to two experimental conditions having networks with shared and differential patterns. We considered three different network structures: (a) Erdos-Renyi networks which randomly generate edges with equal probabilities, (b) small-world networks generated under the Watts-Strogatz model \citep{Watts1998}, and (c) scale-free networks generated using the preferential attachment model  \citep{Barabasi1999} resulting in a hub network. For each type of network, we obtained an adjacency matrix corresponding to the first experimental condition, and then flipped a proportion of the edges in this adjacency matrix to obtain the second network, adding edges where there were no edges and removing an equal number of edges. The proportion of flipped edges was set to 25\%(low), 50\%(medium), and 75\%(high), which correspond to varying levels of discordance between the experimental conditions. 

After generating the networks, the corresponding precision matrices were constructed as follows. For each edge, we generated the corresponding off-diagonal element from a Uniform(-1,1) distribution and fixed the diagonal elements to be one and the off-diagonals corresponding to absent edges as zero. In order to ensure that the resulting precision matrices were positive definite, we subtracted the minimum of the eigenvalues from each diagonal element of the generated precision matrix. To enable a group level comparison for each scenario, all subjects had the same network across all time points within each experimental condition and the same precision matrices for each network.  

\uline{\noindent \it{Tuning:}}
We used BJNL with 1000 burn-in iterations and 5000 MCMC iterations. We specified the tuning parameters as follows. We chose $\lambda_0=100$ and $\tau_{g,kl}\sim Ga(a_\tau, b_\tau)$ with $a_\tau=0.1$ and $b_\tau=1$ in prior specification (\ref{eq:px}) to enforce a sharp spike at zero and thick tails for the slab component. The stick breaking weights in the mixture distribution in (\ref{eq:stickbreak}) were modeled as $\nu_{g,h}\sim Be(1,M)$, where $M \sim Ga(a_m,b_m)$, and we choose $a_m=1,b_m=1,$ to encourage a small number of edge clusters for a parsimonious representation. We could increase $a_m$ to encourage a larger number of clusters. However, we have observed that varying $a_m$ has a limited effect on the final estimated network, as demonstrated through simulations in Section 2 of the Supplementary Materials. Our experience in extensive numerical studies suggests that the performance of the approach is not overly sensitive to the choice of $\lambda_0$ as long as it is large enough ($>100$); however, extremely large values of $\lambda_0$ can result in numerical instability. Moreover, performance is fairly robust to the choice of the hyperparameters in the prior for the precision parameter of the slab component in (\ref{eq:px}), as long as the ratio $a_{\tau}/b_{\tau}<1$.

The joint graphical lasso depends on two tuning parameters: a lasso penalty and a fused lasso penalty. We searched a 30$\times$30 grid over $[0.01, 0.1]$ for both parameters to find the best combination of tuning parameters using a AIC criteria as recommended in Danaher et. al (2014). The graphical lasso was run independently for each network over a grid of regularization parameter values, and the optimal graph was selected for each network using a BIC criteria as described in Yuan and Lin (2007).

\uline{\noindent \it{Performance metrics:}} We assessed the performance of the three algorithms in terms of the ability to estimate the individual networks, as measured by the area under the receiver operating characteristic (ROC) curve (AUC), the accuracy in estimating the strength of connections, as measured by the $L_1$ error in estimating the precision matrix ($L_1$ error), the power to detect true differential edges as measured via sensitivity (TPR) and control over false positives for differential edges which is computed as 1-specificity (FPR). For all the metrics, we performed pairwise comparisons using Wilcoxon signed rank tests in order to assess whether one approach performed significantly better than the others. For edge detection, point estimates for the penalized networks were obtained by choosing the threshold for the absolute off-diagonal elements as 0.005, while for BJNL we computed thresholds controlling for false discoveries as described in Section 3. 

\subsection{Simulation Results}

\begin{figure}
\begin{center}
\includegraphics[width=0.8\textwidth]{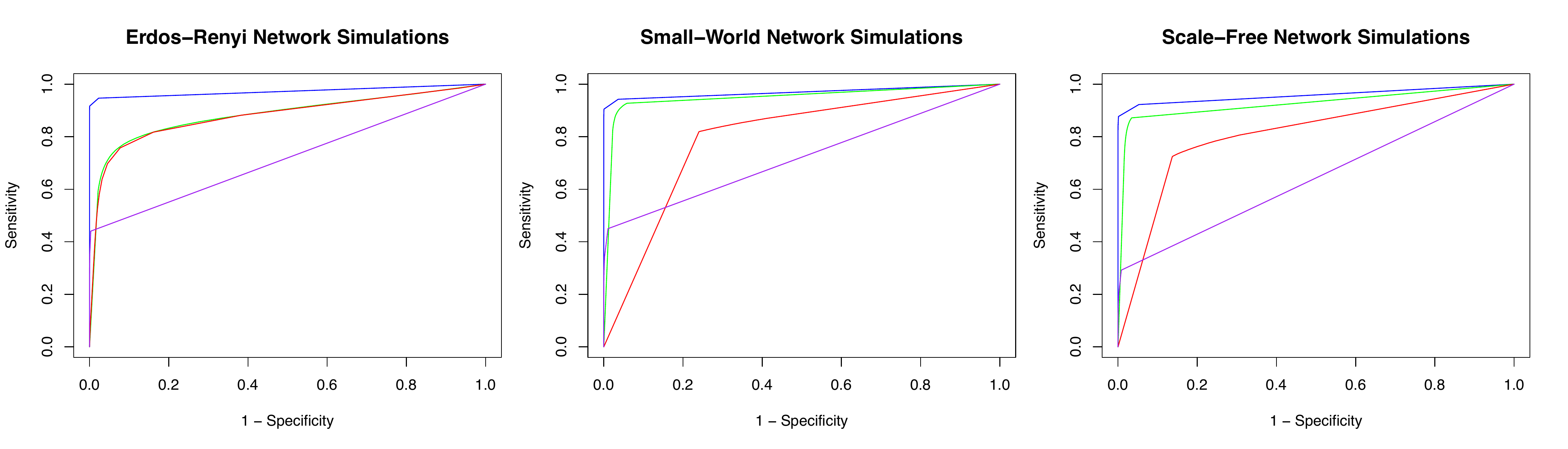}
\caption{\footnotesize ROC curves for edge detection for the 100 node simulations. The blue, green, red, and purple solid lines correspond to BJNL, JGL, GL, and HS respectively.}
\label{ROCsimulation}
\end{center}
\end{figure}

\begin{figure}
	\begin{center}
		\includegraphics[width=6in, height=2.0in]{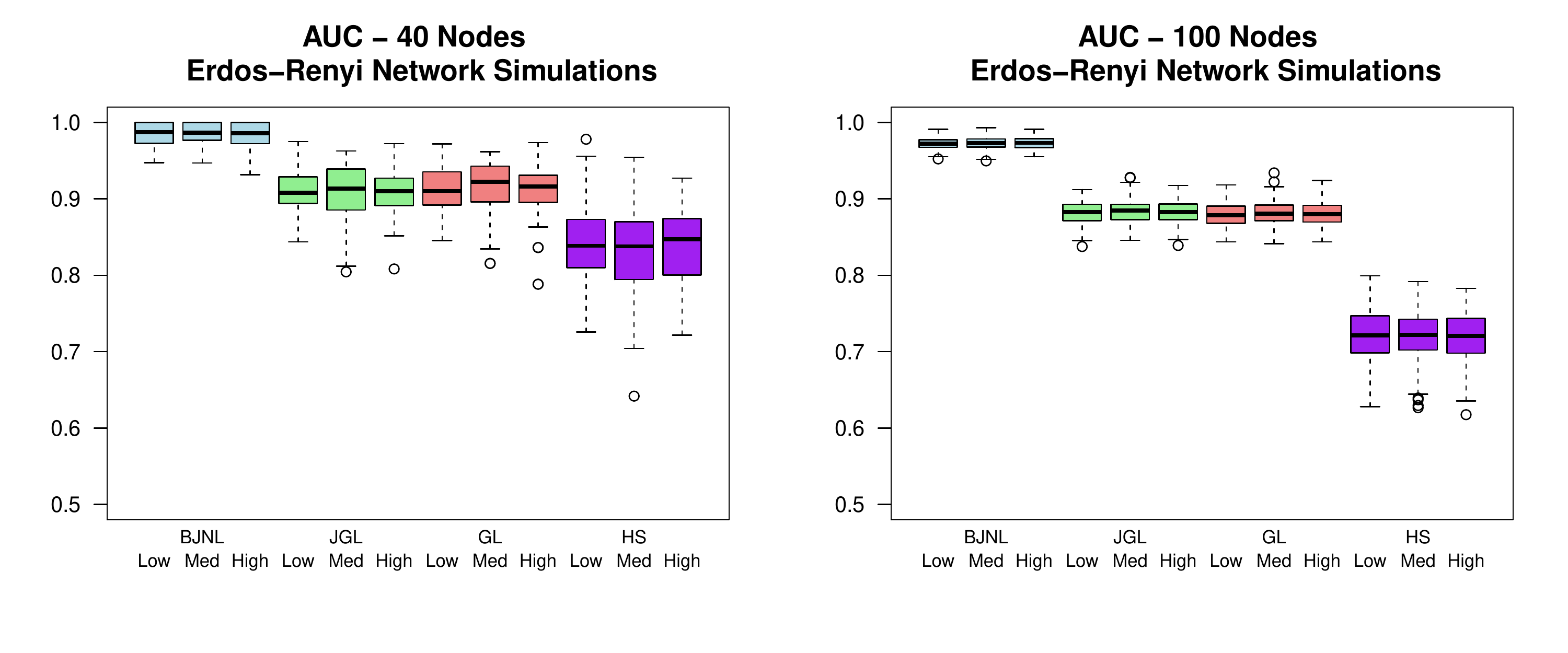} 
		\vspace{-0.1in.} \includegraphics[width=6in, height=2.0in]{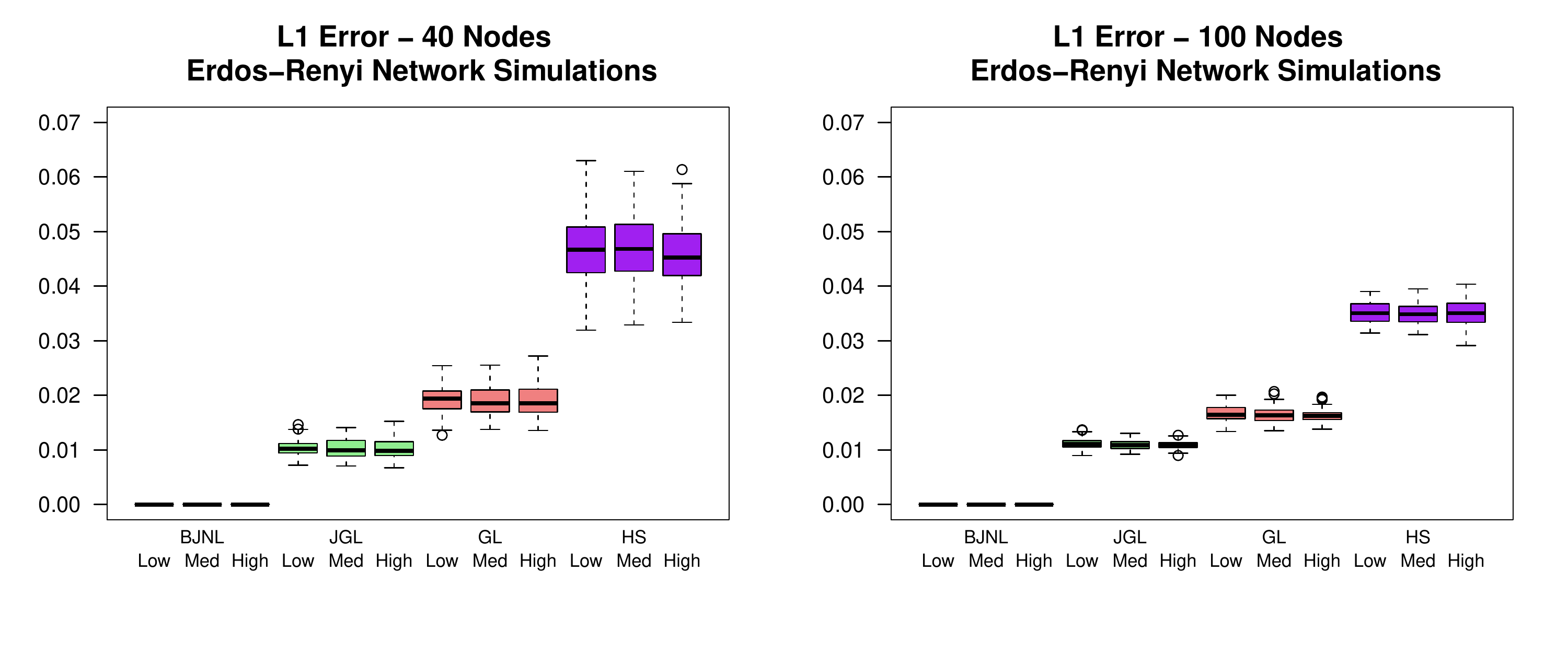} 
		\vspace{-0.1in.}\includegraphics[width=6in, height=2.0in]{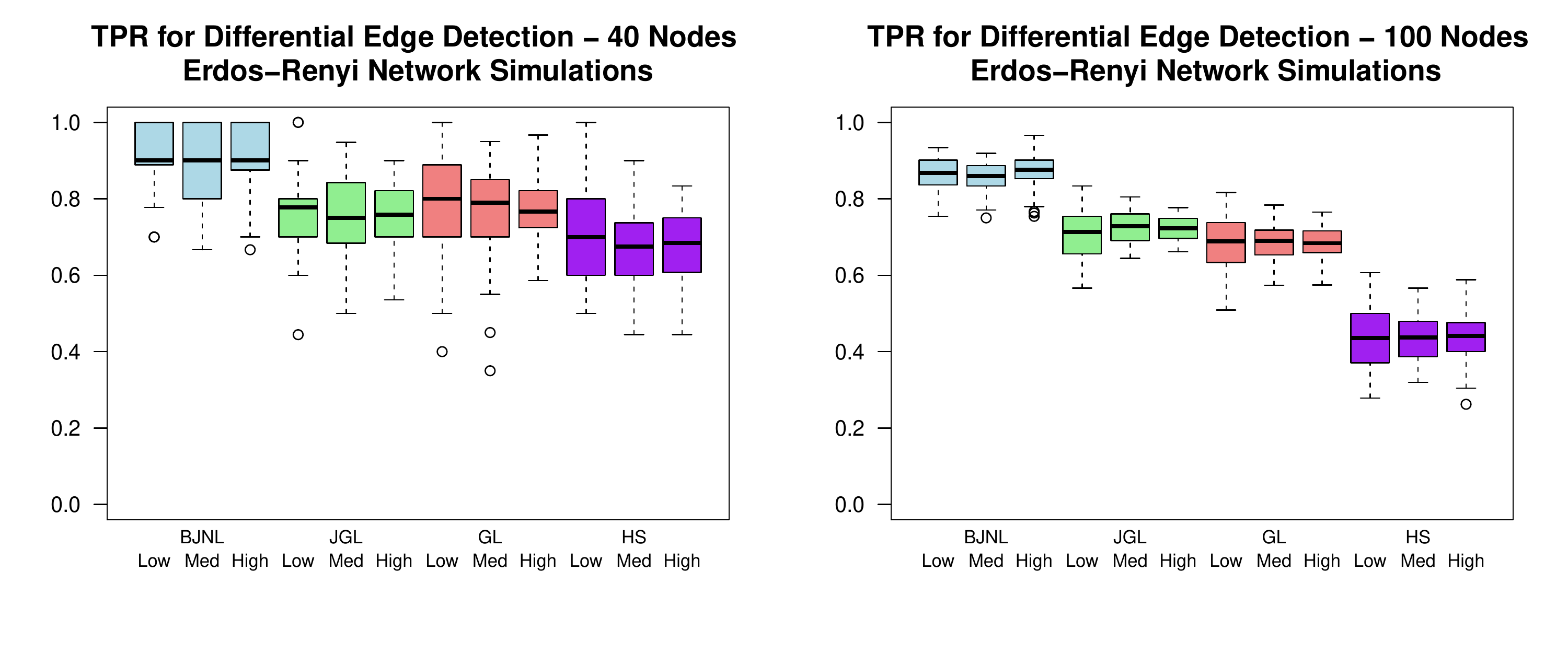}
		\vspace{-0.1in.}\includegraphics[width=6in, height=2.0in]{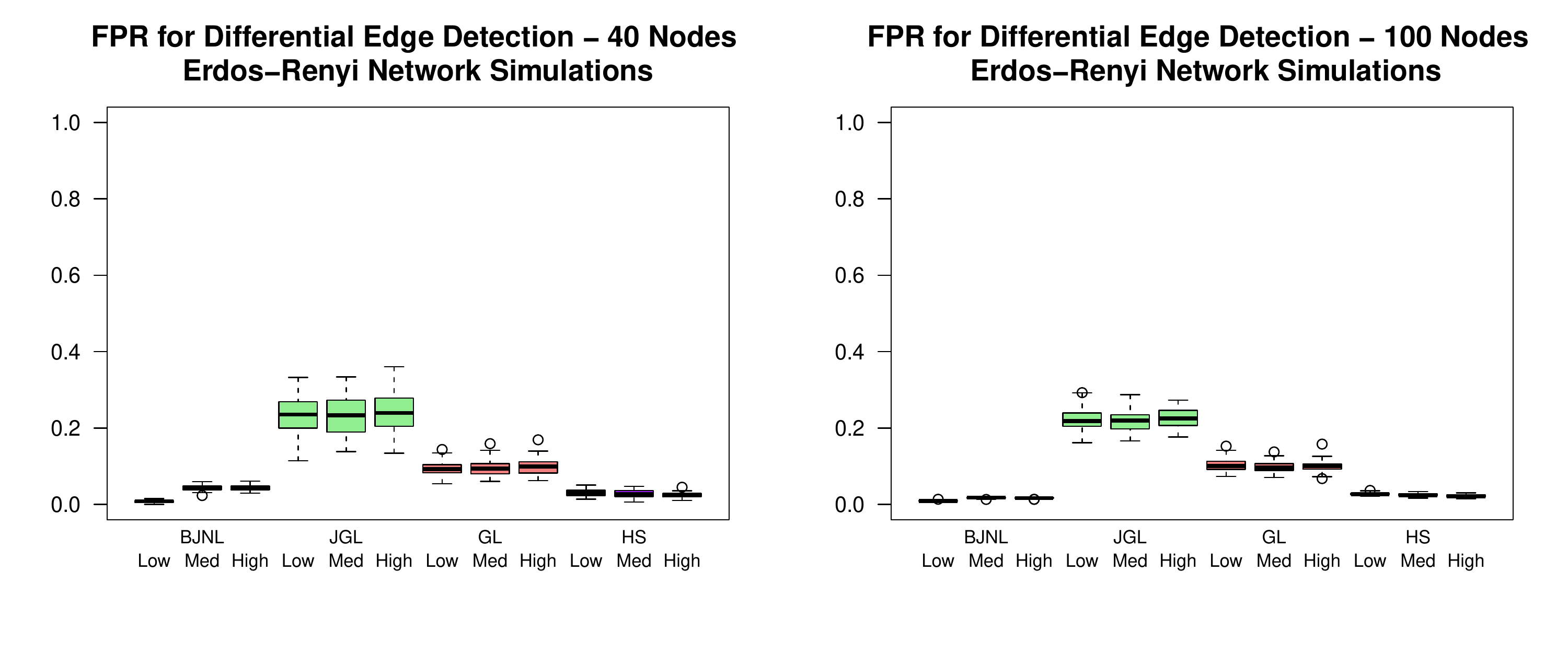}
		\caption{\footnotesize Box plots of the AUC, $L_1$ Error, and TPR/FPR for differential edge detection for the Erdos-Renyi simulations for Bayesian Joint Network Learning (BJNL), the Joint Graphical Lasso (JGL), Graphical Lasso (GL) and the Graphical Horseshoe Estimator (HS). Within each approach, the box plots are organized as: low difference, medium difference, and high difference in edges between experimental conditions, in that order.}
		\label{ERresults}
	\end{center}
\end{figure}

Figure \ref{ROCsimulation} displays the ROC curves for the 100 node simulations, Figure \ref{ERresults} displays box plots of the reported metrics for the Erdos-Renyi case, and Table \ref{simulationResults} reports results for the 100 node simulations. The box plots for the other networks and the results for the 40 node case are reported in the Supplementary Materials due to space constraints. The results across the three network types are relatively consistent. First, we note that the degree of dissimilarity between the networks does not appear to have a major effect on the relative performance of the algorithms, although we conjecture that the differences could be more pronounced for smaller sample sizes. For all settings involving Erdos-Renyi graphs, the proposed BJNL approach outperformed the HS, JGL, and GL uniformly across all metrics under the Wilcoxon signed rank test. Notably, the proposed approach simultaneously achieved a significantly higher TPR and a significantly lower FPR for differential edges, indicating that it was both better able to detect significant differences and less likely to incorrectly classify an edge as differential. These, and the additional box plots in the Supplementary Materials, suggest a greater power to detect true differential edges  with an adequate control over false positives across all network types, under the BJNL. Further, an increased improvement of the TPR over competing approaches and relative stability of the FPR  for differential edges for $p=100$ versus $p=40$  indicates a clear advantage of the proposed joint estimation approach for increasing dimensions. For the small-world and scale-free networks, the BJNL also had significantly improved AUC, TPR, and $L_1$ error metrics, and a comparable or lower FPR, compared to all other considered approaches. 


On the  other hand, the significantly higher $L_1$ error under the JGL potentially points to the perils of smoothing over edge strengths across networks under penalized approaches. In particular, assigning similar magnitudes for precision matrix off-diagonals for shared edges may adversely affect the identification of differential edges, as well as the estimation of varying edge strengths for common edges across networks. Moreover while HS has low FPR, it consistently exhibits the lowest AUC and TPR and the highest $L_1$ error for $p=100$ across all scenarios, which is concerning. On the other hand, the GL had the highest FPR for both the small-world and scale-free network simulations, but has a reasonable TPR. These results under HS and GL illustrate the difficulties resulting from the separate estimation of individual networks which may result in exceedingly low power to detect true positives (as with HS), or an inflated number of false positives (as with GL).

To examine the sensitivity of the proposed approach with respect to the chosen link function, we performed additional simulation studies by fitting the proposed model to the 100 node data generated as above, but under a probit link. The results in Table \ref{probitTable} illustrate non-significant differences in the performance metrics for network estimation across the logit and the probit links, which illustrate the robustness of the proposed approach resulting from the specification of the DP prior on the shared and differential components in (\ref{eq:weights}).

\begin{table}[]
	\centering
	\caption{\small 100 node simulation results comparing Bayesian Joint Network Learning (BJNL), the Joint Graphical Lasso (JGL), Graphical Lasso (GL) and the Graphical Horseshoe Estimator (HS). Text in bold indicates a method was better than all other competing methods as assessed through Wilcoxon signed rank tests at $\alpha = 0.05$.}
	\label{simulationResults}
	\begin{adjustbox}{angle=0, width=0.85\textwidth, height=0.75\textheight, keepaspectratio}
\begin{tabular}{|c|cccc|cccc|} \hline
	& \multicolumn{4}{c|}{AUC}                          & \multicolumn{4}{c|}{$L_1$ Error $\times$ 100}                 \\ 
	 & BJNL                 & JGL         & GL    &     HS      & BJNL                 & JGL                  & GL & HS          \\ \hline
	 	Erdos-Renyi & & & & & & & & \\
	low         & \textbf{0.97 (0.01)} & 0.88 (0.02) & 0.88 (0.02) & 0.72 (0.03) & \textbf{0.11 (0.01)} & 1.11 (0.09)          & 1.66 (0.13) & 3.51 (0.19)\\
	med         & \textbf{0.97 (0.01)} & 0.88 (0.02) & 0.88 (0.02) & 0.72 (0.04) & \textbf{0.11 (0.01)} & 1.09 (0.09)          & 1.65 (0.14) & 3.50 (0.20) \\
	high        & \textbf{0.97 (0.01)} & 0.88 (0.02) & 0.88 (0.02) & 0.73 (0.03) &\textbf{0.11 (0.01)	} & 1.09 (0.07)          & 1.62 (0.11) & 3.50 (0.23) \\
	Small World &                      &             &             &          &            &                      &    &         \\
	low         & \textbf{0.97 (0.01)} & 0.95 (0.01) & 0.79 (0.01) &  0.72 (0.04) & \textbf{0.25 (0.01)} & 0.75 (0.12)          & 2.06 (0.08) & 4.70 (0.15) \\
	med         & \textbf{0.97 (0.01)} & 0.95 (0.01) & 0.80 (0.01) & 0.72 (0.03)& \textbf{0.24 (0.01)} & 0.77 (0.13)          & 2.07 (0.08) & 4.65 (0.14) \\
	high        & \textbf{0.97 (0.01)} & 0.95 (0.01) & 0.79 (0.01) & 0.73 (0.03) & \textbf{0.24 (0.01)} & 0.78 (0.13)          & 2.06 (0.08) & 4.65 (0.14) \\
	Scale Free  &                      &             &             &                   &   &              &        &             \\
	low         & \textbf{0.96 (0.01)} & 0.93 (0.01) & 0.81 (0.01) & 0.64 (0.03) &\textbf{0.20 (0.01)} & 1.01 (0.20)          & 2.23 (0.10) & 5.30 (0.23) \\
	med         & \textbf{0.96 (0.01)} & 0.92 (0.01) & 0.81 (0.01) & 0.64 (0.03) &\textbf{0.19 (0.01)} & 1.02 (0.21)          & 2.24 (0.90) & 5.26 (0.24) \\
	high        & \textbf{0.96 (0.01)} & 0.92 (0.01) & 0.81 (0.01) & 0.64 (0.03) &\textbf{0.19 (0.01)} & 1.00 (0.21)          & 2.20 (0.08) & 5.23 (0.23) \\ \hline
	& \multicolumn{4}{c|}{TPR}                          & \multicolumn{4}{c|}{FPR}                                    \\ 
	 & BJNL                 & JGL         & GL    & HS      & BJNL                 & JGL                  & GL   & HS       \\ \hline
	Erdos-Renyi & & & & & & & &\\
	low & \textbf{0.87 (0.05)} & 0.71 (0.07) & 0.68 (0.07) & 0.43 (0.08) & {0.01 (0.001)} & 0.22 (0.03)   & 0.10 (0.02) & 0.03(0.00) \\
	med         & \textbf{0.88 (0.04)} & 0.73 (0.04) & 0.69 (0.05) & 0.44 (0.06) & {0.01 (0.001)} & 0.22 (0.03)          & 0.10 (0.01) & 0.03(0.00) \\
	high        & \textbf{0.88 (0.02)} & 0.72 (0.03) & 0.69 (0.04) & 0.44 (0.06) &{0.01 (0.001)} & 0.23 (0.02)          & 0.10 (0.02) & 0.02 (0.00) \\
	Small World &                      &             &             &            &          &                 &     &             \\
	low         & \textbf{0.86 (0.04)} & 0.47 (0.07) & 0.66 (0.06) & 0.44 (0.07) & 0.02 (0.002)         & 0.02 (0.00) & 0.36 (0.01) & 0.06 (0.01) \\
	med         & \textbf{0.86 (0.04)} & 0.49 (0.04) & 0.67 (0.04) & 0.46 (0.05) &  0.02 (0.002)          & 0.02 (0.00) & 0.36 (0.01) & 0.05 (0.01) \\
	high        & \textbf{0.86 (0.02)} & 0.48 (0.04) & 0.67 (0.03) & 0.46 (0.05) & 0.01 (0.002)          & 0.02 (0.00) & 0.36 (0.01) & 0.05 (0.01) \\
	Scale Free  &                      &             &             &                      &            &          &      &       \\
	low         & \textbf{0.87 (0.05)} & 0.39 (0.06) & 0.63 (0.07) & 0.25 (0.06) & 0.02 (0.002)          & 0.02 (0.00) & 0.24 (0.03) & 0.04 (0.01) \\
	med         & \textbf{0.87 (0.03)} & 0.41 (0.05) & 0.63 (0.04) & 0.26 (0.05) &0.02 (0.002)        & 0.02 (0.00) & 0.24 (0.02) & 0.04 (0.01) \\
	high        & \textbf{0.87 (0.03)} & 0.42 (0.04) & 0.64 (0.04) & 0.27 (0.05) & 0.01 (0.002)         & 0.02 (0.00) & 0.25 (0.02) & 0.04 (0.01) \\ \hline
\end{tabular}
\end{adjustbox}

\caption{Comparison of the 100 node simulation results using the probit link function to the simulation results using the logit link function. }
\label{probitTable}
\begin{adjustbox}{angle=0, width=0.5\textwidth, height=0.5\textheight, keepaspectratio}
\begin{tabular}{|lccc|ccc|ccc|}
\hline
& \multicolumn{3}{c|}{Erdos Renyi} &  \multicolumn{3}{c|}{Small World} &  \multicolumn{3}{c|}{Scale Free} \\  \hline
& AUC & TPR & FPR & AUC & TPR & FPR & AUC & TPR & FPR\\
Probit & 0.97 & 0.88 & 0.01 & 0.96 & 0.86 & 0.02 & 0.97 & 0.87 & 0.02\\
Logit  & 0.97 & 0.88 & 0.01 & 0.97 & 0.87 & 0.02 & 0.96 & 0.86 & 0.02\\ \hline
\end{tabular}
\end{adjustbox}

\end{table}

\section{Stroop task analysis}

\subsection{Description of Analysis}

We applied the proposed BJNL to the fMRI Stroop task study to investigate similarities and differences in the brain network under the two experimental conditions and passive fixation (REST). The first analysis was aimed at comparing the mental states of task performance (TASK) and passive fixation (REST), with the hypothesis that the brain networks exhibit major differences between these two grossly different conditions. The TASK data consisted of the subject-wise concatenation of the prewhitened fMRI time courses acquired during the EXR and RLX blocks, while the REST data consisted of the subject-wise concatenation of the prewhitened fMRI time courses acquired during the passive fixation blocks. The second analysis aimed to detect finer differences in connectivity between the mental states of EXR and RLX task performance. The study hypothesized that the mental states should be similar between the two task conditions with some fine differences in the network. In this case, the subject-wise prewhitened fMRI time courses were concatenated for the EXR blocks and also separately for the RLX blocks, corresponding to the two experimental conditions to be compared.

We performed a brain network analysis based on region of interest (ROI) level data, adopting the 90 node Automated Anatomical Labeling (AAL) cortical parcellation scheme described in \cite{Tzourio-Mazoyer2002}. For each ROI, we estimated the representative BOLD time series by performing a singular value decomposition on the time series of the voxels within the ROI and extracting the first principal time series. This resulted in 90 time courses of fMRI measurements, one for each ROI, which were then demeaned. We classified each ROI into one of nine functional modules as defined in \cite{Smith2009} using the technique described in \cite{Kemmer2015}. We performed standard pre-processing including slice-timing correction, warping to standard Talairach space, blurring, demeaning, and pre-whitening. The fMRI time series was prewhitened using an AR(1) model, as is common in imaging toolboxes such as AFNI \citep{cox1996} and SPM \citep{penny2011}. Further details are provided in Section 5 of the Supplementary Materials. The proposed BJNL was run using the same tuning parameters as in the simulations. Dickey-Fuller tests of stationarity were performed to assess convergence of the MCMC sampler (see Section 6 of the Supplementary Materials). We also examined the widths of the credible intervals in Section 7 of the Supplementary Materials, where Figure 7 of the Supplementary Materials demonstrates that the credible intervals for absent-edges are much narrower than the credible intervals for present-edges. Finally, we performed chi-squared and related goodness of fit tests to verify that the BJNL approach provides an adequate fit to the Stroop task data (see Section 8 of the Supplementary Materials).

\uline{\noindent \it{Graph metrics:}} We analyzed the brain's connectivity structure during the different mental states in terms of four graph metrics: global efficiency, local efficiency, clustering coefficient, and characteristic path length. Efficiency measures how effectively information is transmitted from node-to-node in a network. Global efficiency measures information transmission across the entire graph and is calculated by taking the average across all ROIs of the inverse shortest path lengths between ROIs. Thus, large values of global efficiency indicate that, on average, the number of steps required to transmit information from one node to another is small. Local efficiency measures information transmission between an ROI and its neighbors and is calculated for each ROI by taking the average of the inverse shortest path lengths between ROIs in the relevant neighborhood, where the relevant neighborhood is the collection of ROIs with a connection to the selected ROI. The clustering coefficient measures the interconnectedness of the graph and is calculated for each ROI by examining how many of its neighbors are also neighbors to each other. Finally, characteristic path length is the average across ROIs of the shortest path length in the networks, with smaller values indicating a more efficient network. All graph metrics were calculated using the Matlab Brain Connectivity Toolbox \citep{Rubinov2010}. Differences in the graph metrics values across mental states were computed at each MCMC iteration, and the central tendency and dispersion of their distributions were statistically assessed by T-tests and Kolmogorov-Smirnov tests.

\subsection{Results}
\begin{figure}
	\begin{center}
		\includegraphics[width=0.75\linewidth]{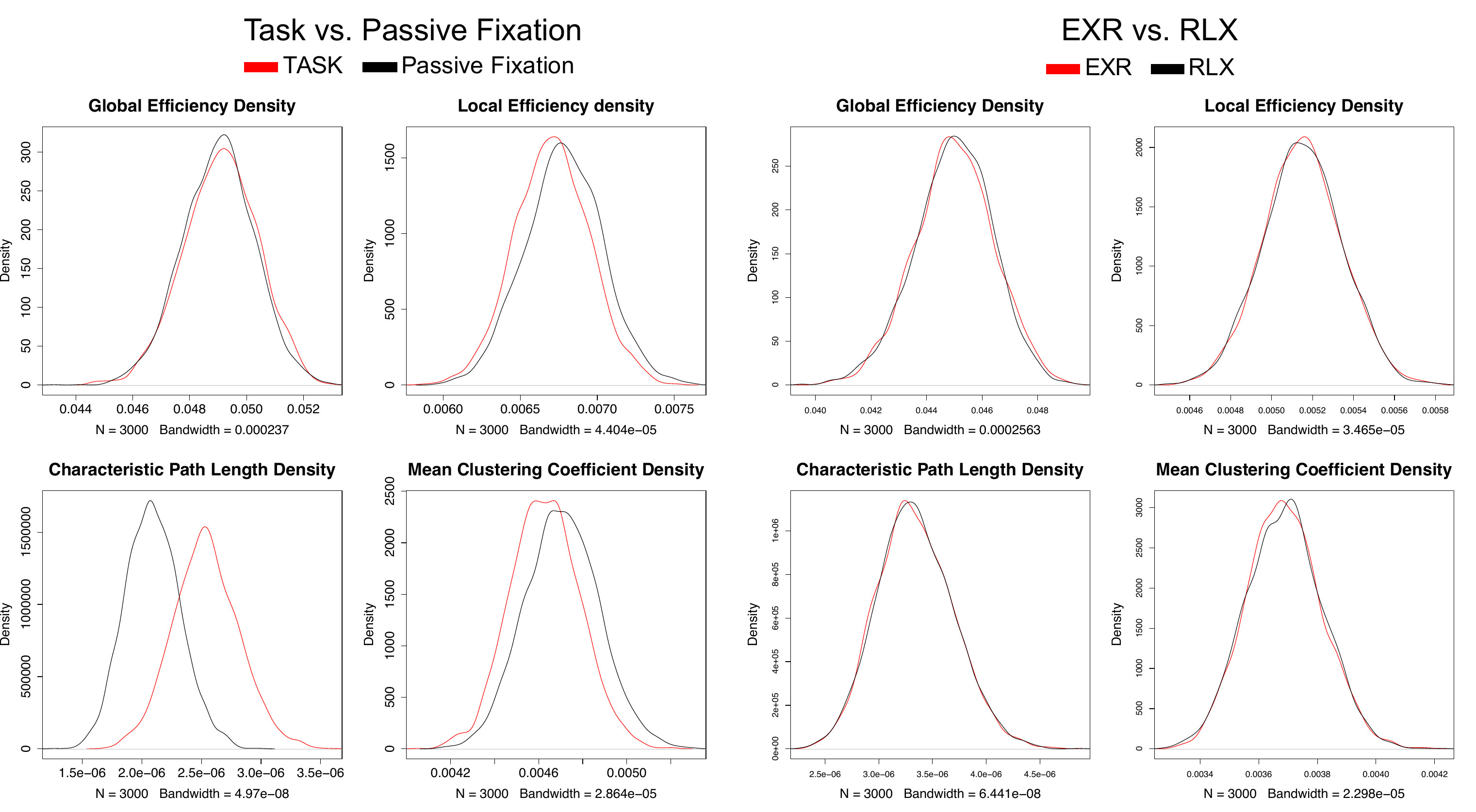} 
		\caption{\footnotesize Estimated densities of graph metrics for the analysis of task vs. passive fixation and maximum exertion (EXR) vs. relaxed (RLX) task performance.}
		\label{graphMetrics}
	\end{center}
\end{figure}

\begin{figure}
	\begin{center}
		\includegraphics[height=2in, width=0.9\linewidth]{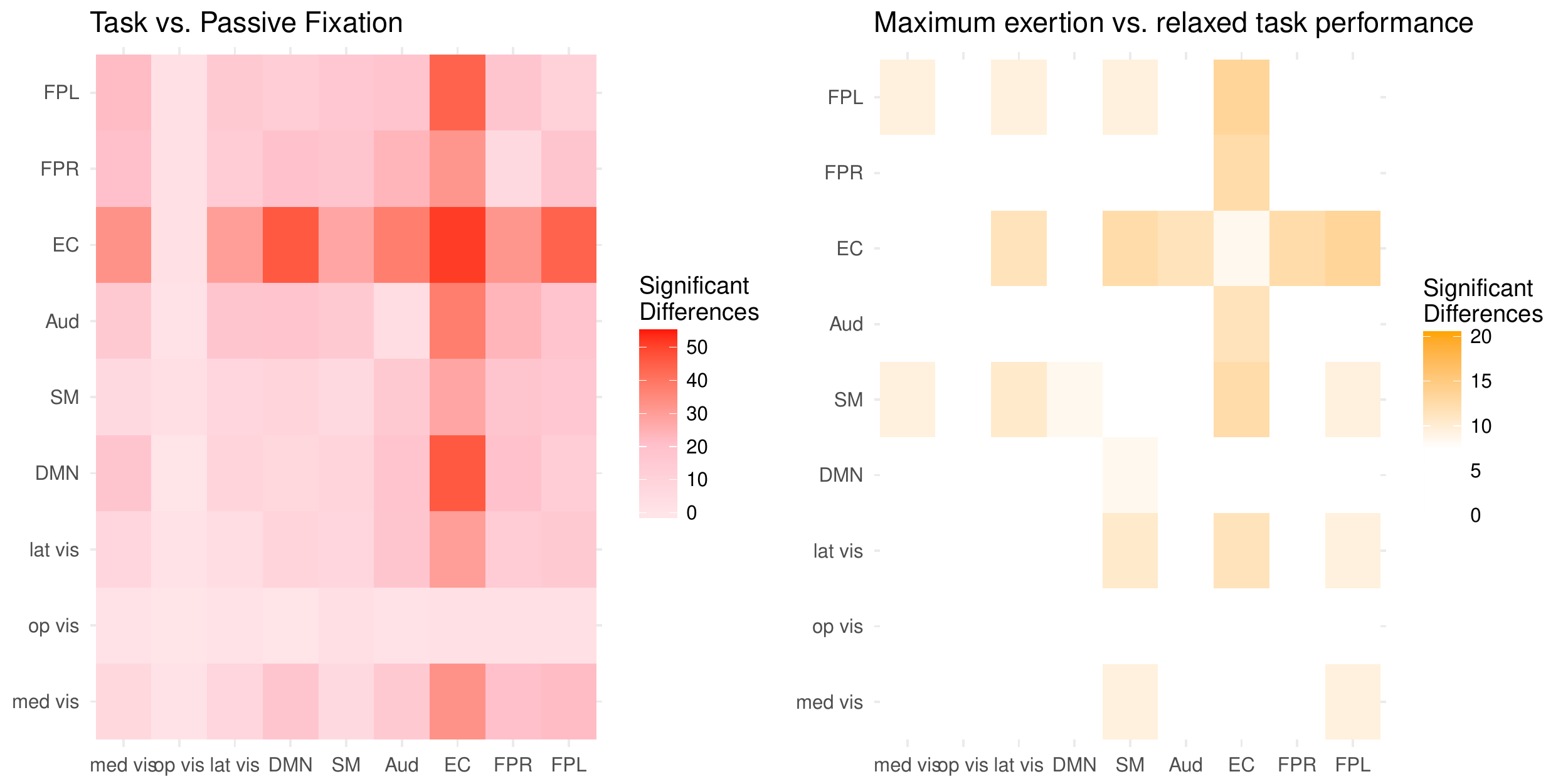}
		\caption{\footnotesize Heatmaps of the number of differential edges between conditions. The heatmap on the left corresponds to the analysis of task vs. passive fixation, and the heatmap on the right corresponds to the analysis of maximum exertion (EXR) vs. relaxed task performance (RLX).}
		\label{differenceHeatmaps}
	\end{center}
\end{figure}

{\noindent \uline{TASK vs REST Conditions:}} The analysis revealed a large contingent of edges with significantly different strengths in the two mental states. These results provide evidence supporting the study hypothesis that there are major differences in the brain networks due to the manifest phenomenological and procedural dissimilarity of task performance and rest. T-tests (p $<$ 0.01, FDR-corrected) of the Fisher Z-transformed  partial correlation differences were conducted as a post-processing step using all the MCMC samples after burn-in. This analysis revealed 763 significantly different edges, with 618 out of the 763 edges lying within our adopted functional module partition.  Figure \ref{differenceHeatmaps} displays a heatmap of the significant edge counts by functional module.  Moreover, our examination of network metrics revealed significant differences in the mean and the posterior distributions for all network metrics between the two conditions (Figure \ref{graphMetrics}). 

{\noindent \uline{EXR vs. RLX conditions of task performance:}} Compared with the relatively large network differences between TASK and REST, the network structures corresponding to the two task conditions, i.e. EXR and RLX, were much more similar to each other, as hypothesized by the investigator. T-tests  (p $<$ 0.01, FDR-corrected) of the Fisher's Z-transformed partial correlation differences were conducted as a post-processing step using all the MCMC samples after burn-in. This analysis revealed 247 significantly different edges between the EXR and RLX conditions, 226 of which lay within our functional module partition. Figure \ref{differenceHeatmaps} displays a heatmap of the significant edge counts by functional module. Also, none of the differences between the graph metrics for the two states were significant, providing evidence that the overall network features such as efficiency and path lengths between the two states are quite similar (Figure \ref{graphMetrics}). 

{\uline{Interpretation of Findings:}} The above analyses provide an application of BJNL to both grossly and subtly different cognitive conditions. As expected, we identified drastic differences between Stroop task performance and passive fixation (REST). Descriptively, the TASK condition was associated with stronger positive connections involving frontoparietal circuits, DMN, sensorimotor, and visual cortices compared to the passive fixation condition. We also found significant differences in all graph metrics we calculated, which suggests highly different patterns of information transmission between task performance and passive fixation. These differences outline a picture of an overall state of more efficient brain connectivity during REST. For example the characteristic path length was significantly smaller during passive fixation than during task performance, indicating that information transmission between any two nodes requires fewer steps. This finding provides exciting new insight into the neuro-physiological differences between the brain's intrinsic network architecture under the REST condition and the task-related network in which the brain requires modification of some connections in order to perform the task.

The analysis of EXR vs. RLX revealed fewer dissimilarities, as expected due to the only difference between conditions being in the executive stance with which the subjects were instructed to perform the task. We found no global topological differences between the EXR and RLX conditions. However, BJNL did reveal some fine differences in the functional modules including the EC and FPL. These networks are involved in high level cognitive function. Notably, the within-EC differences are much less pronounced in the analysis of EXR vs. RLX than TASK vs. REST. In general, the relaxed task performance condition featured significantly more negative connections between regions.

As an illustrative alternative analysis, we used a permutation testing procedure to assess the differences in edge strengths. The labels for EXR and RLX were randomly permuted, and the JGL and GL were used to estimate two separate graphs. We then saved the edge-wise difference in the graphs for each permutation. This procedure was repeated 10,000 times, and empirical $p$-values were calculated for the difference in edge strengths under the true labels. The p-values were then FDR-corrected, as with the BJNL analysis. None of the resulting p-values were significant for the GL, and only 2 of the p-values were significant for the JGL. Similarly, for the analysis of TASK vs REST, the JGL identified no differential edges and 624 common edges (versus 763 differential edges and 1211 common edges under BJNL). In this case, the GL was able to identify 136 edges with differential strengths (27 of which overlap with those identified by the BJNL), and 661 common edges. We believe that the absence of differential edges between EXR versus RLX conditions under the GL, and between TASK versus REST conditions under JGL, is unrealistic for the Stroop task experiment. These results suggests the proposed BNJL method has much better statistical power to detect differences in brain networks under different cognitive states compared to an approach involving independent estimation of networks.

\section{Discussion}

In this paper we introduced a novel Bayesian approach to joint estimation of multiple group level brain networks that pools information across networks to estimate shared and differential patterns. To our knowledge, ours is one of the first Bayesian approaches to address this important problem. Through a wide variety of simulation studies, we demonstrated  clear and significant advantages of the proposed joint estimation approach over commonly used penalized approaches, with such improvements becoming more pronounced as the number of nodes increases. The method was applied to a Stroop task dataset, and the analysis revealed important dissimilarities between the task and rest conditions, but more subdued differences between the two task conditions. In contrast, alternate analyses using penalized approaches were not able to identify nearly as many differential edges.

Although the proposed BJNL typically results in more accurate estimation of multiple group level networks and provides measures of uncertainty, it is slower than the penalized methods and the total computation time increases with the number of experimental conditions and the number of nodes. However, the latter is true of any graphical modeling approach. It is important to note that the proposed method can be made considerably faster by adopting a parallel computation scheme which samples the $G$ precision matrices in parallel given the networks $\mathcal{G}_1,\ldots,\mathcal{G}_G$. Future work should investigate the scalability of BJNL to larger numbers of conditions while taking into account the dynamic nature of the brain networks over time. In this paper, we demonstrated BJNL for estimating networks using fMRI data because they are the most prevalent type of functional images. However, our method can also be generalized to data from other imaging modalities in a straightforward manner. One advantage of our proposed approach for clustering the edge weights is that it allows for unsupervised estimation of the number of clusters. This means that in generalizing the method to other modalities, we do not have to laboriously tune the clustering parameters to each individual problem. Going beyond multiple experimental conditions, our approach can also be used to jointly model networks across multiple cohorts, such as healthy individuals, subjects with mild cognitive disorder, and those with Alzheimer's disease. 

\section{Supplementary Materials}
The Supplementary Materials contain the posterior computation steps in detail, the results of the 40 node simulation studies, the boxplots for performance metrics under the small world and scale-free networks simulation scenarios, and additional details on the Stroop Task data analysis. \phantom{\cite{cox1996, dickey1979, yuan2007}}

\bibliography{BJGLreferences.bib}

\end{document}